\documentclass{article}
\usepackage{ace,amsmath,graphicx,url,times,mathtools,multirow,balance}
\usepackage[utf8]{inputenc}
\usepackage{cmdtrack}

\newcommand{\rtsixty}[1]{{$\mbox{RT}_{60}$}}

\pdfoutput=1
\title{SRMR VARIANTS FOR IMPROVED BLIND ROOM ACOUSTICS CHARACTERIZATION}

\name{Mohammed Senoussaoui, Jo\~ao F. Santos and Tiago H. Falk}
\address{INRS-EMT, University of Quebec, Montreal, QC, Canada}

\begin{document}

\ninept
\maketitle


\begin{abstract}

Reverberation, especially in large rooms, severely degrades speech recognition performance and speech intelligibility. Since direct measurement of room characteristics is usually not possible, blind estimation of reverberation-related metrics such as the reverberation time (RT) and the direct-to-reverberant energy ratio (DRR) can be valuable information to speech recognition and enhancement algorithms operating in enclosed environments. The objective of this work is to evaluate the performance of five variants of blind RT and DRR estimators based on a modulation spectrum representation of reverberant speech with single- and multi-channel speech data. These models are all based on variants of the so-called Speech-to-Reverberation Modulation Energy Ratio (SRMR). We show that these measures outperform a state-of-the-art baseline based on maximum-likelihood estimation of sound decay rates in terms of root-mean square error (RMSE), as well as Pearson correlation. Compared to the baseline, the best proposed measure, called NSRMR$^*_k$, achieves a 23\% relative improvement in terms of RMSE and allows for relative correlation improvements ranging from 13\% to 47\% for RT prediction.
\end{abstract}


\section{Introduction}
\label{sec:intro}

In an enclosed environment, a speech signal recorded by a far-field microphone is often affected by reverberation, which is the addition of multiple attenuated reflections of the source signal. Reverberant speech leads to a severe degradation in performance of automatic speech recognition systems, as well as lower intelligibility. This effect is highly dependent on the room characteristics, and quantified objectively by measures that depend on the room impulse response (RIR). One such metric is the reverberation time (RT), which is the required time for the sound energy to decay by a certain amount (e.g., 60 dB, which is denoted \rtsixty~) after the extinction of the sound source \cite{sabine1922collected}. The direct-to-reverberant energy ratio (DRR) is another well-known measure related to reverberation, and consists of the energy ratio between the energy of the sound coming directly from the source and its reflections \cite{JoDRR}. Measuring the RIR, however, is not always possible, especially in real-time applications. As such, several blind room acoustics characterization methods (i.e., methods that estimate room characteristics from reverberant speech only) have been proposed in the literature (e.g., \cite{Ratnam_2003,4517613,Loll_IWAENC_2010,Prego}), with more recent ones relying on modulation spectral information extracted from the reverberant speech signal (e.g., \cite{5422672,5547575}).

The modulation spectrum represents the temporal dynamics of the envelopes of frequency subbands of a speech signal. Due to articulation characteristics, most of the energy in speech is concentrated in low modulation frequencies (i.e., in the 2 - 20 Hz range). The addition of multiple reflections, however, generates higher frequency envelope modulations. As such, previous works have shown that the relationship between low and high frequency envelope modulations contains relevant information related to the reverberant environment and the so-called Speech-to-Reverberation Modulation Energy Ratio (SRMR) was developed \cite{5422672,5547575}. In \cite{5547575}, the SRMR metric was used to predict speech quality and intelligibility of reverberant and dereverberated speech; in \cite{5422672}, a variant of the metric was used to predict \rtsixty~ and DRR. Notwithstanding, SRMR was recently shown to result in high inter- and intra-speaker variability \cite{6309396} and a normalization procedure was developed for intelligibility prediction \cite{IWAENC_JOAO}.

In this paper, we investigate the performance of the normalized SRMR metric within the scope of the Acoustic Characterization of Environments (ACE) Challenge \cite{ACE2015}, as well as several other variants as correlates of \rtsixty~ and DRR. The variants are proposed to i) maintain temporal per-frame modulation spectral cues for SRMR estimation (as opposed to the use of an average modulation spectrum, as in the original SRMR formulation), ii) normalize the modulation spectrum to reduce inter- and intra-speaker variability, and iii) perform multi-channel analyses. Performances of the proposed \rtsixty~ metrics are compared to the maximum-likelihood estimation of sound decay rates metric proposed in \cite{Loll_IWAENC_2010}. Experimental results show several of the proposed SRMR variants outperforming the benchmark algorithm across several testing conditions.


\section{Modulation spectral representations of reverberant speech}
\label{sec:ReverberantSR }

\begin{figure}[t]
  \centering
  \centerline{\includegraphics[width=8.5cm,height=3.1cm]{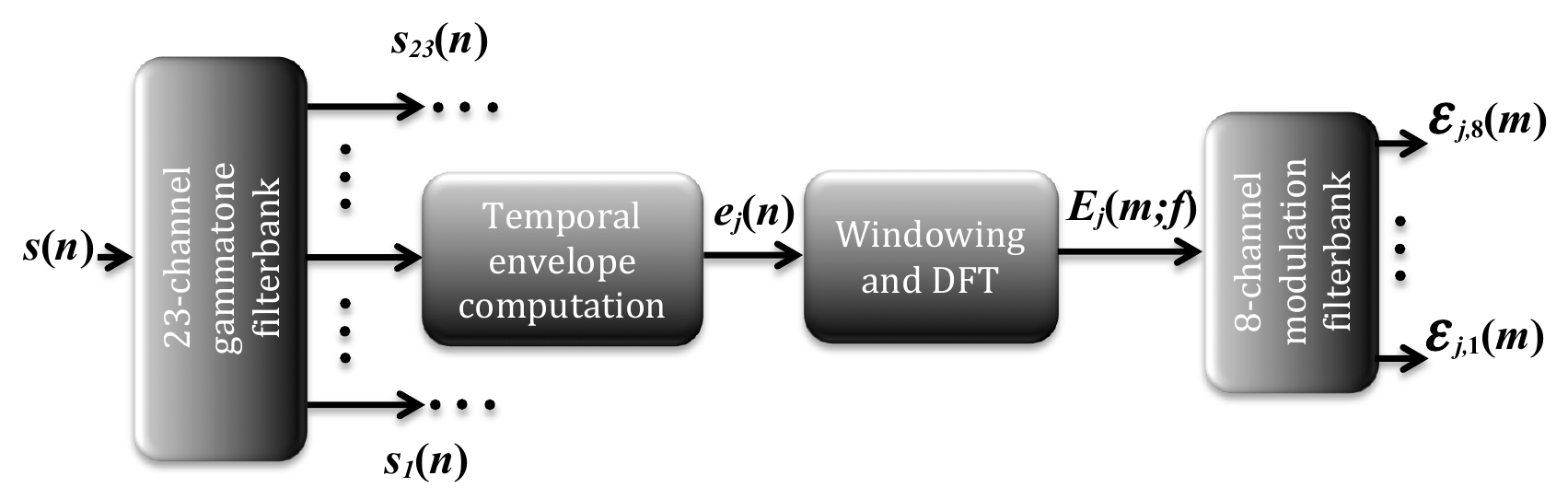}}
  \caption{Block diagram of modulation spectral processing steps.}
  \label{fig:MSprocess}
\end{figure}

Figure~\ref{fig:MSprocess} summarizes the signal processing steps used to extract the modulation spectral representation used in the computation of the SRMR metric. For a given input speech signal $s(n)$, a critical-band gammatone filterbank, with 23 filters, is first applied in order to emulate the human cochlea \cite{317247}. Second, a Hilbert transform $\mathcal{H}\{.\}$ is used to capture the temporal dynamics information from the output of each gammatone filter $s_j(n)$ where $j \in {[1,\dots,23]}$. Next, each temporal envelope $e_j(n)$ is segmented by means of a 256-ms Hamming sliding window with 32-ms shifts. For each frame $m$, the corresponding windowed envelope $e_j(m)$ is then subject to the discrete Fourier transform $\mathcal{F}\{.\}$ to obtain the modulation spectrum for critical-band $j$ denoted as $E_j(m,f)$, where $f$ is the modulation frequency. Lastly, the modulation frequency bins are grouped into K-bands in order to emulate an auditory-inspired modulation filterbank. The $k^{th}$ modulation band energy for the $m^{th}$ frame and the $j^{th}$ gammatone filter is denoted as $\varepsilon_{j,k}(m)$ and it represents one entree of a tensor $\varepsilon$ of dimension $23 \times 8 \times M$, where $M$ indexes the number of frames extracted. More details about the extraction process can be found in \cite{5422672}. Based on this $23 \times 8 \times M$ representation, two metrics were previously proposed:

\subsection{Per-modulation band SRMR: SRMR$_k$}
In \cite{5422672}, the original per-modulation band SRMR was proposed, also known as SRMR$_k$. This metric was given by
\begin{equation} \label{eq:SRMR_k}
\mbox{SRMR}_k=\frac{\displaystyle\sum_{j=1}^{23}\displaystyle\sum_{m=1}^{M}\varepsilon_{j,1}(m)}{\displaystyle\sum_{j=1}^{23}\displaystyle\sum_{m=1}^{M}\varepsilon_{j,k}(m)},
\end{equation}
where $k$ indicated the index of the modulation filter used and ranged from $5 - 8$. For the \rtsixty~ prediction task, the four dimensional vector comprised of $[\mbox{SRMR}_5, \mbox{SRMR}_6, \mbox{SRMR}_7, \mbox{SRMR}_8]$ was used as feature to a support vector regression (SVR) model.

\subsection{Overall SRMR: OSRMR}
In \cite{5422672}, an overall SRMR (OSRMR) metric was proposed and shown to be highly correlated with DRR. The metric was computed as:
\begin{equation} \label{eq:OSRMR}
\mbox{OSRMR}=\frac{\displaystyle\sum_{j=1}^{23}\displaystyle\sum_{m=1}^{M}\varepsilon_{j,1}(m)}{\displaystyle\sum_{k=5}^{8}\displaystyle\sum_{j=1}^{23}\displaystyle\sum_{m=1}^{M}\varepsilon_{j,k}(m)}.
\end{equation}

\subsection{SRMR}
The metrics described above were originally proposed for the purpose of blind room acoustics characterization and relied on the hypothesis that the four last modulation bands (i.e., $k=5-8$) conveyed information about the reverberation tail and the first modulation band ($k=1$) conveyed information about the direct path signal. For speech quality assessment, on the other hand, the SRMR metric was updated such that the ``speech component" incorporated information from the first four modulation bands ($k=1-4$), as opposed to just the first, i.e.,:
\begin{equation} \label{eq:SRMR}
\mbox{SRMR}=\frac{\displaystyle\sum_{k=1}^{4} \displaystyle\sum_{j=1}^{23}\displaystyle\sum_{m=1}^{M}\varepsilon_{j,k}(m)}{\displaystyle\sum_{k=5}^{8}\displaystyle\sum_{j=1}^{23}\displaystyle\sum_{m=1}^{M}\varepsilon_{j,k}(m)}.
\end{equation} 
Here, SRMR is tested as a correlate of \rtsixty~.

\subsection{Normalized SRMR: NSRMR}
In order to reduce the inter- and intra-speaker variability of the SRMR metric, two normalization steps were recently introduced in \cite{IWAENC_JOAO}. First, the modulation frequency ranges used by original SRMR were reduced to alleviate the effects of pitch smearing into the modulation representation. For the purpose of quality assessment, the optimal modulation frequency range was found to be $4 - 40$ Hz. Second, to reduce speech content effects, the dynamic range of the modulation energies was limited to 30dB of the peak average energy. These normalization steps resulted in a relative reduction in root mean square estimator error (RMSE) of approximately 40\% relative to the original SRMR \cite{IWAENC_JOAO} for the task of speech intelligibility prediction. Henceforth, the normalized metrics will be referred to as NSRMR (\rtsixty~ correlate) and NOSRMR (DRR correlate).

\subsection{Per acoustic band SRMR: SRMR$^*_k$}
As can be seen from (\ref{eq:SRMR}), the original SRMR metric was computed based on marginalization of the modulation spectrum over three different dimensions. The first marginalization was over time ($M$ frames), then over the gammatone channels, and lastly over modulation channels. Here, we propose a reformulation of the SRMR metric in which the effect of reverberation to each frame is computed on a per acoustic band basis, thus marginalization is first performed over modulation bands. Marginalization over gammatone filterbank channels and time are given as a final step in the new per acoustic band SRMR (SRMR$^*_k$):
\begin{equation} \label{eq:NovelSRMR}
\mbox{SRMR}^*_k=\frac{1}{M}\displaystyle\sum_{j=1}^{23}\displaystyle\sum_{m=1}^{M}\frac{\varepsilon_{j,1}(m)}{\varepsilon_{j,k}(m)},
\end{equation}
where $k=5$ was chosen empirically. It is important to emphasize that the per-band SRMR metric was computed using the normalized modulation spectrum, thus the terminology $\mbox{NSRMR}^*_5$ will be used. Here, $\mbox{NSRMR}^*_5$ is tested as a correlate of both \rtsixty~ and DRR.

\section{Experimental setup}
\label{sec:Experiments}
\subsection{Dataset description}
Within the context of the ACE challenge, participants were provided with two different datasets, namely, the development and evaluation sets. Challenge participants that required training were invited to use their own datasets. In total, the development set contained 1675 utterances representing the following configurations: five different microphone configurations (1, 2, 3, 5, 8 and 32 channels), three noise types (ambient, babble and fan), and three signal-to-noise ratio (SNR) levels (0dB, 10dB and 20dB). The evaluation set consisted of 4500 utterances per microphone configuration (same as in the development set) under the same noise type conditions but slightly different SNR levels (-1dB, 12dB and 18dB).

\begin{figure}[t]
  \centering
  \centerline{\includegraphics[width=8.7cm,height=6.3cm]{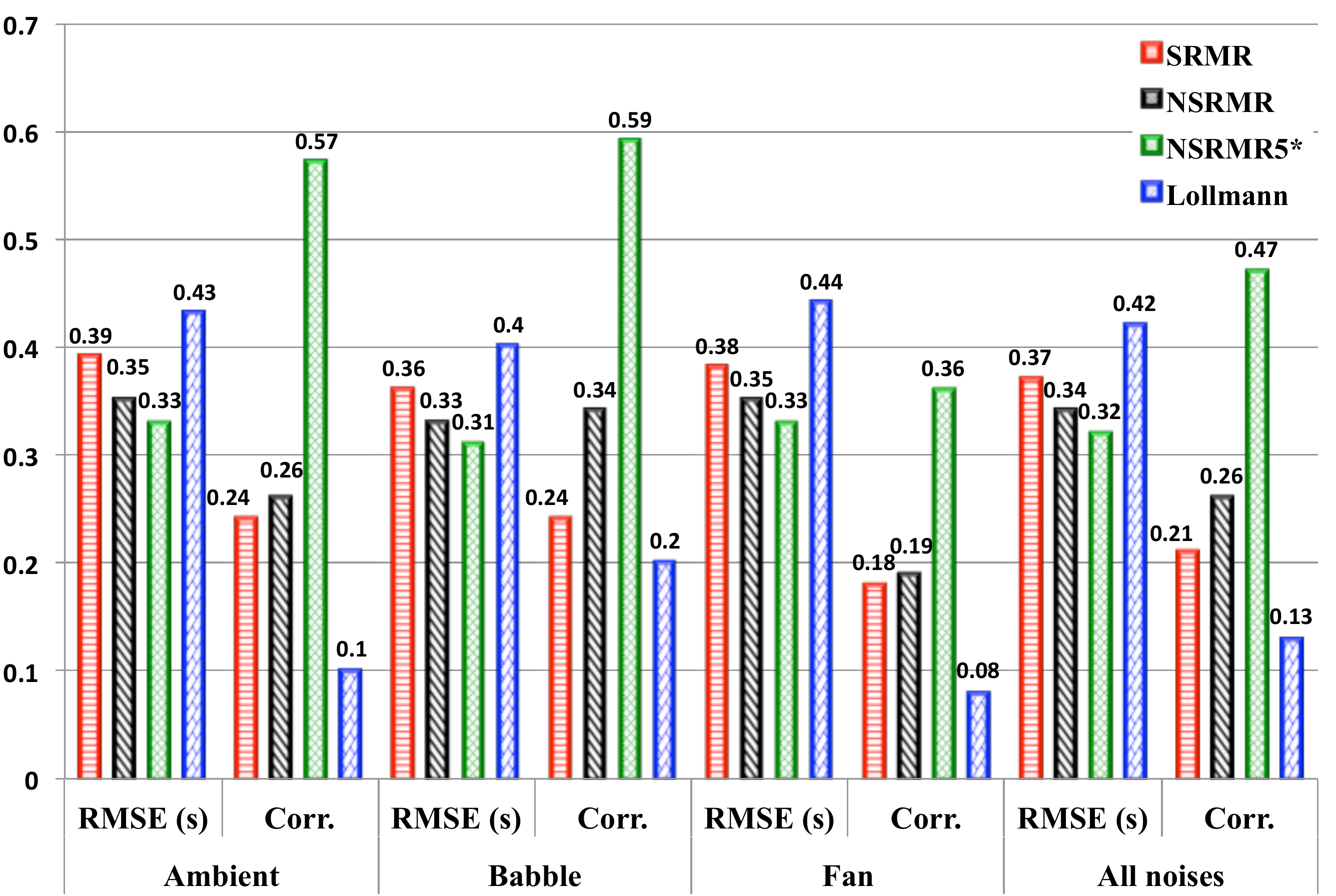}}
  \caption{Single-channel RMSE (in seconds) and correlation of \rtsixty~ predictors.}
  \label{fig:RT60_barPlot_Single}
\end{figure}

For the proposed metrics, new mappings needed to be obtained from the SRMR variants to their corresponding \rtsixty~ or DRR values. As such, we utilized the test part of the widely-used TIMIT database convolved with artificial or recorded RIRs, and with additive noise. For simulated RIRs, the image method was used \cite{allen79,Peterson} and reverberant speech with \rtsixty~ values ranging from $0.25-1.05$ s (with $0.2$ s increments) were synthesized. The recorded RIRs, in turn, were taken from the Aachen Impulse Response (AIR)  database described in detail in \cite{Jeub_DoWeNeed,Jeub_binaural}. The database is comprised of RIRs collected in four rooms with varying microphone-speaker distances; a range of approximately $RT_{60}=0.2-0.8$ s is available. Lastly, the reverberant speech signals were further corrupted by additive noise using two noise types (metro station and restaurant) taken from the Diverse Environments Multichannel Acoustic Noise Database (DEMAND)\footnote{http://parole.loria.fr/DEMAND/} at SNR levels ranging from 0-20dB in 10dB increments. 
\subsection{Parameter mapping}
The ACE Challenge was comprised of four tasks. The first two were dedicated to predicting fullband \rtsixty~ or DRR parameters, whereas the second two to predicting the parameters per 1/3-octave ISO subbands. In this work, focus was placed only in the two fullband tasks. As such, fullband DRR and \rtsixty~ values were computed from the artificial and recorded RIRs and used as ground truth to train the mappings. In our experiments, a linear regression mapping was trained for the DRR estimators and a generalized linear regression model (GLM) based on a normal distribution and logarithmic link function configuration was used for \rtsixty~ prediction. 
Moreover, the level of the reverberant speech signal was first normalized to -26 dB overload (dBov) using the ITU-T P.56 voltmeter \cite{ITU-T-P.56} prior to SRMR feature extraction. 
\subsection{Multi-channel analyses}
The SRMR metric was originally proposed for single-channel data and the variants described here were developed for single microphones. Nonetheless, two multi-microphone strategies were explored. The first considered each channel in a multi-channel setup separately and finally averaged the multiple estimated parameters into a final value. The second approach comprised averaging the per-channel SRMR metric and its variants over all the channels prior to mapping. In the development set, the latter approach resulted in improved performance. As such, the results reported herein utilize feature averaging over multiple channels as a simple strategy for multi-channel blind room acoustics characterization. In our experiments, only the 2-, 3-, and 5-channel cases were used.
\subsection{Figures of merit and benchmark algorithm}
For the \rtsixty~ estimators, two performance parameters are used as figures of merit: the Pearson correlation between the estimated and true parameters, as well as the root mean square error (RMSE) expressed in seconds. In order to gauge the benefits of the proposed estimators, the Maximum Likelihood based method described in \cite{Loll_IWAENC_2010} is used as benchmark; henceforth, the method will be referred to as Lollman's method. The relative gains obtained in correlation and RMSE with the proposed metrics over the benchmark are reported. For the multi-channel cases, the average of the benchmark outputs was used for comparisons. On the other hand, to the best of the authors knowledge, there are no published blind DRR estimators, thus a DRR benchmark is not available. For DRR estimators only the RMSE (expressed in decibels) is used as figure of merit.

\begin{table*}[t]
{\centering
{
\begin{tabular}{l|c|c|c|c|}
\cline{2-5}
                                            & \multicolumn{1}{l|}{{\bf Single (1-Ch.)}} & \multicolumn{1}{l|}{{\bf Chromebook (2-Ch.)}} & \multicolumn{1}{l|}{{\bf Mobile (3-Ch.)}} & \multicolumn{1}{l|}{{\bf Crucif (5-Ch.)}} \\ \hline\hline
\multicolumn{1}{|l|}{{ SRMR (\rtsixty~)}}   						& 0.12                                      & 0.12                                          & 0.10                                     	& 0.11                                      \\ \hline
\multicolumn{1}{|l|}{{ NSRMR (\rtsixty~)}}    						& 0.12                                      & 0.12                                          & 0.09                                      	& 0.10                                       \\ \hline
\multicolumn{1}{|l|}{{ $\mbox{NSRMR}^*_5$  (\rtsixty~)}} 			& 0.10                                      & N/A                                             	& N/A                                         	& N/A                                         \\ \hline
\multicolumn{1}{|l|}{{ Lollmann (\rtsixty~)}}   						& 0.18                                      & 0.15                                          & 0.14                                      	& 0.15                                      \\ \hline\hline\hline
\multicolumn{1}{|l|}{{ OSRMR (DRR)}}     						& 16.58                                    & {\bf6.87}                                   & 11.45                                     	& 13.98                                     \\ \hline
\multicolumn{1}{|l|}{{ NOSRMR (DRR)}}    						& 16.60                                    & {\bf8.28}                                   & 12.07                                    	& 14.70                                      \\ \hline
\multicolumn{1}{|l|}{{ $\mbox{NSRMR}^*_5$ (DRR)}}  				& 21.96                                    & N/A                                            	& N/A                                         	& N/A                                         \\ \hline
\end{tabular}
\caption{Prediction error variance for single- and multi-channel estimation scenarios.}
\label{Tab:Var}}}
\end{table*}

 \begin{figure}[t]
  \centering
  \centerline{\includegraphics[width=8.7cm,height=5.3cm]{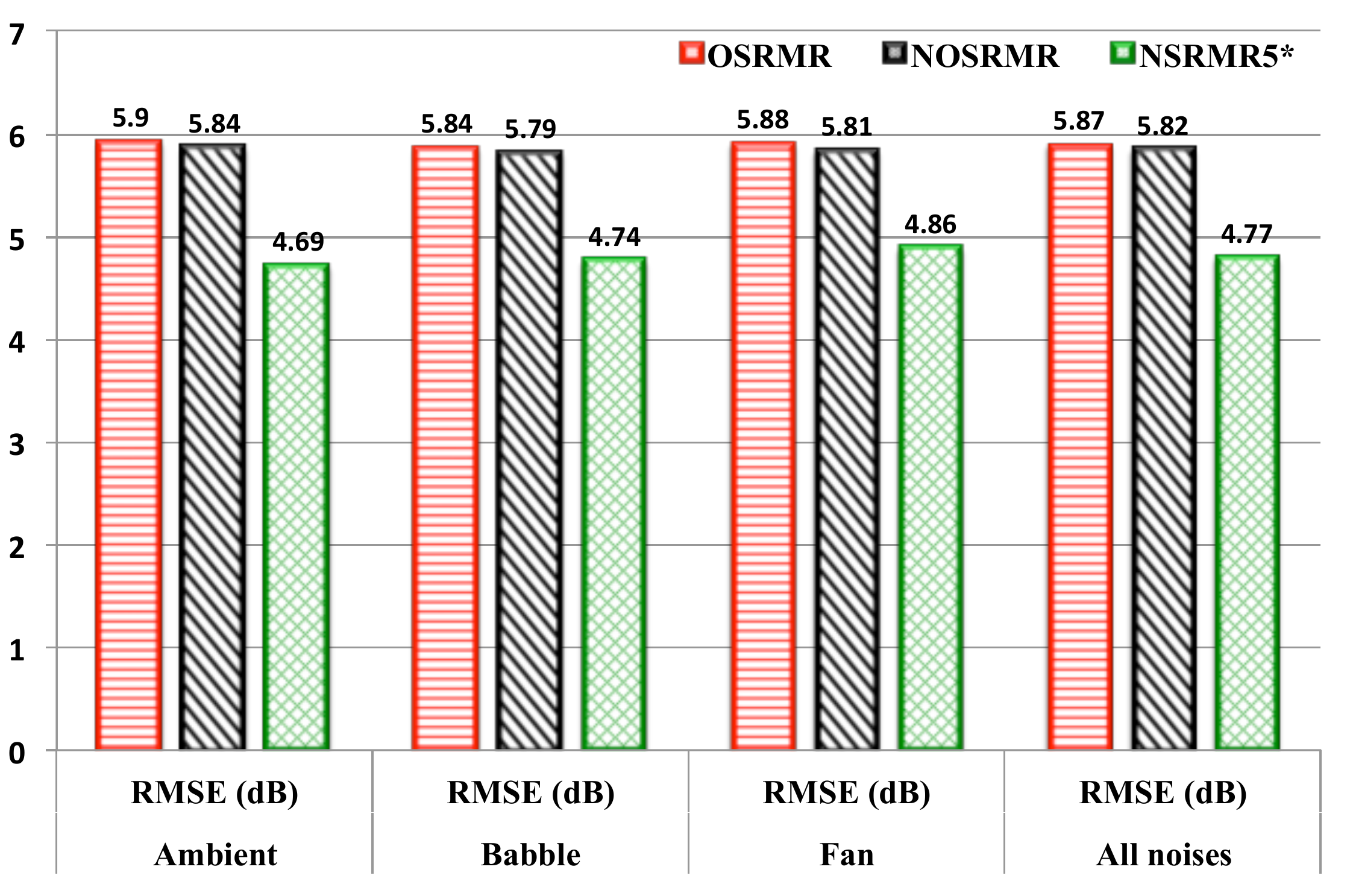}}
  \caption{Single-channel RMSE (in dB) of DRR predictors.}
  \label{fig:DRR_barPlot_Single}
\end{figure}

\section{Experimental Results}
\subsection{Single channel}
\label{sssec:ResDiscusT60SingCh}
Figure~\ref{fig:RT60_barPlot_Single} shows the two figures of merit for the single-channel \rtsixty~ estimates obtained with SRMR, NSRMR, $\mbox{NSRMR}^*_5$, and Lollman's method. Results are reported by acoustic noise type (i.e. Ambient, Babble, Fan) as well as with all noise types combined (in both cases, averaged over all noise levels). As can be seen, in terms of RMSE, the proposed $\mbox{NSRMR}^*_5$ metric achieves results inline with NSRMR, but requires information from only the first and fifth modulation bands, as opposed to all eight modulation bands used in NSRMR. Both variants, in turn, outperform the original SRMR. Overall, all SRMR variants outperformed the benchmark and showed small variability as a function of noise type. Table~\ref{Tab:Var} (first column) reports the overall variance of the estimation error for the different \rtsixty~ estimators for the single-channel case.

In terms of correlation, however, the proposed $\mbox{NSRMR}^*_5$ metric achieved significantly higher results relative to other SRMR-based parameters and the benchmark method. Correlation values were close to 0.6 for ambient and babble noise conditions, but only 0.36 in the fan noise condition. In the latter case, despite the low correlation attained, the obtained results were still significantly better than the benchmark, which achieved a 0.08 correlation coefficient, thus exemplifying the difficulty of the task in fan-noise conditions. Overall, relative improvements in correlation with the proposed $\mbox{NSRMR}^*_5$ metric ranged from 13\% to 47\% for the \rtsixty~ estimation over the benchmark. Moreover, Fig.~\ref{fig:DRR_barPlot_Single} depicts the RMSE attained with the DRR estimators. Comparing Fig.~\ref{fig:DRR_barPlot_Single} and Table~\ref{Tab:Var}, it can be seen that while $\mbox{NSRMR}^*_5$ outperformed OSRMR and NOSRMR in terms of RMSE (by as much as 18\% across all noise types), it achieved a higher prediction error variance, thus suggesting the parameter was less stable in its estimates.

\begin{figure}[t!]
  \centering
  \centerline{\includegraphics[width=8.7cm,height=5.3cm]{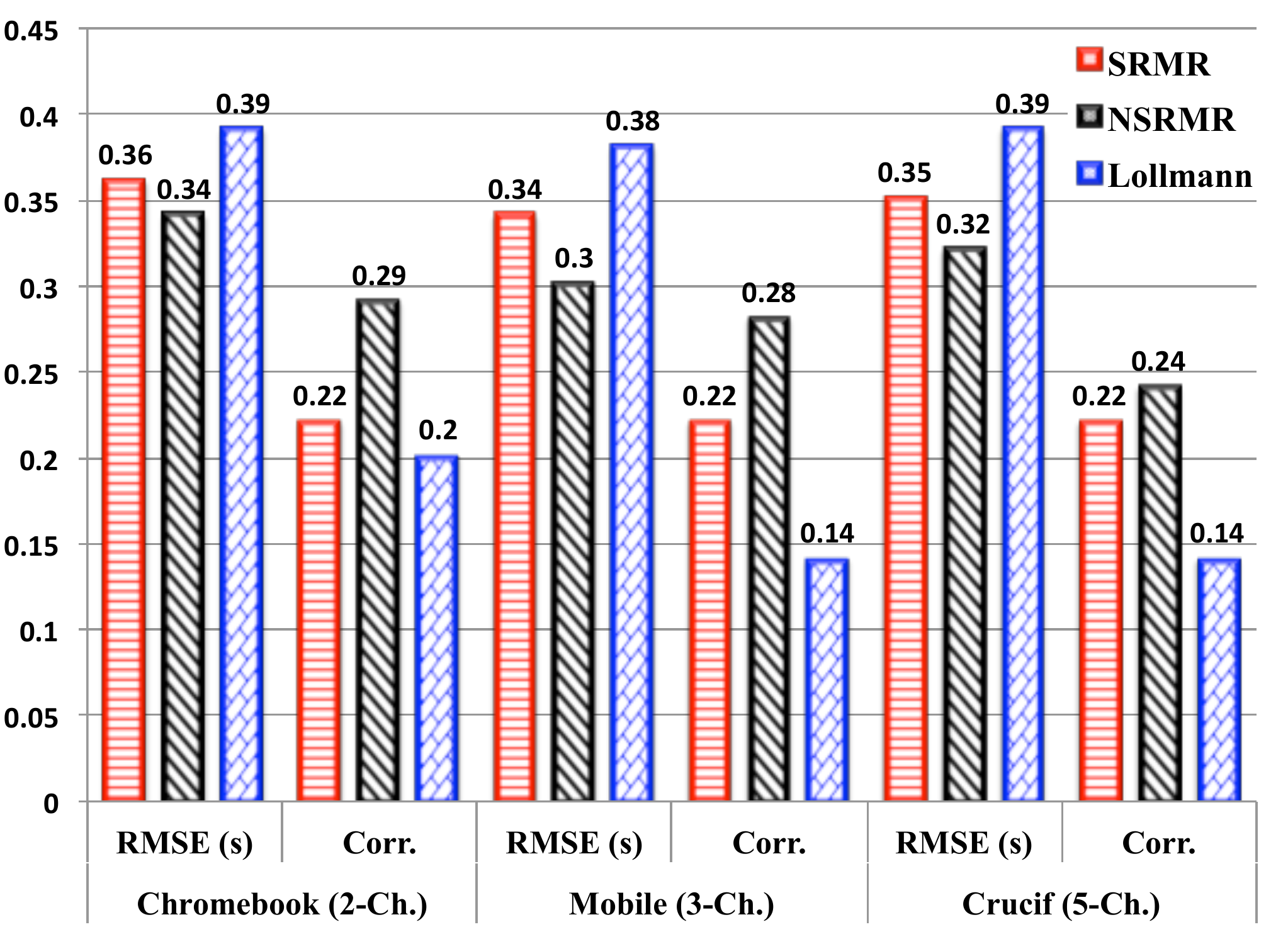}}
   \caption{Multi-channel RMSE (in seconds) and correlation of \rtsixty~ predictors.}
  \label{fig:T60_barPlot_Multi}
\end{figure}

\subsection{Multi-channel}
\label{sssec:ResDiscusMultiCh}
Figure~\ref{fig:T60_barPlot_Multi} shows the two figures of merit for the multi-channel \rtsixty~ estimators. Due to limited challenge submissions, prediction data is not available for the $\mbox{NSRMR}^*_5$ estimator in multi-channel settings. Notwithstanding, the achieved RMSE with the SRMR and NSRMR metrics are in-line with those achieved in the single-channel case, with NSRMR achieving slightly lower RMSE values. Both predictors achieved RMSE lower than the Lollman benchmark method. From Table~\ref{Tab:Var} (columns 2-4), it can also be observed that the prediction error variance remained comparable to those obtained in the single-channel case with a slight decrease as the number of channels increased; both proposed metrics also outperformed the benchmark.

\begin{figure}[t!]
  \centering
  \centerline{\includegraphics[width=8.5cm,height=5.3cm]{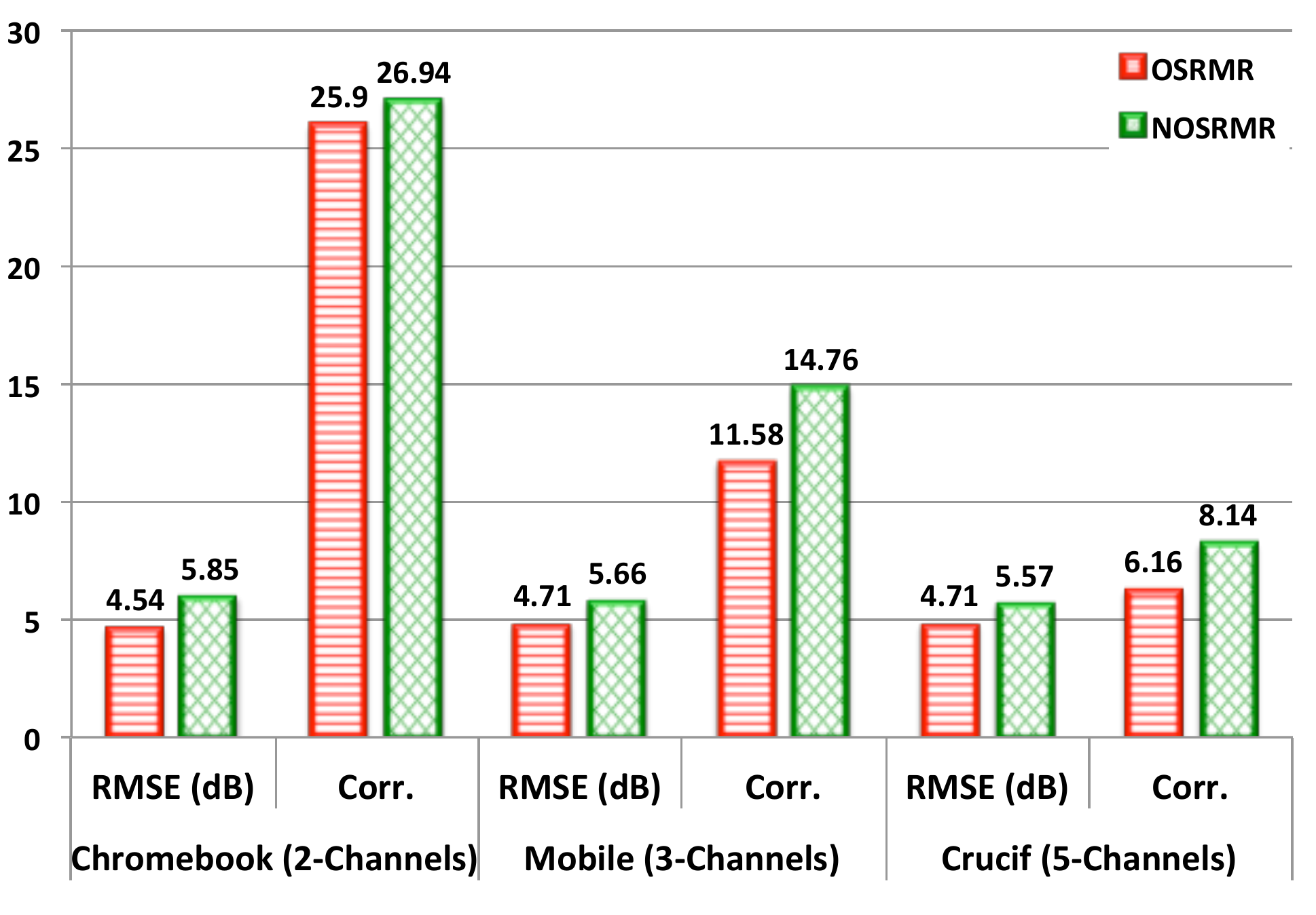}}
 \caption{Multi-channel RMSE (in dB) of DRR predictors.}
  \label{fig:DRR_barPlot_Multi}
\end{figure}

Similar findings were observed with the correlation parameter. Correlation values comparable to those achieved in the single-channel scenario were obtained. Unfortunately, the proposed $\mbox{NSRMR}^*_5$ metric which showed significant gains in correlation in the single-channel case was not available in the multi-channel case. This investigation is left for a future study. Lastly, Fig. \ref{fig:DRR_barPlot_Multi} shows the RMSE of the DRR predictions. As can be seen, the performance of the OSRMR metric is slightly better than NOSRMR in the multi-channels case. From Table~\ref{Tab:Var} it can be seen that the same is true for the prediction error variability. Interestingly, the prediction error variability was the lowest in the 2-channel case with a drop of over 50\% relative to the single-channel case. Overall, the multi-channel results suggest that further improvements are likely possible from more complex multi-channel grouping strategies. This is the focus of our ongoing work.

\section{Conclusion}
\label{sec:Conclusion}
In this work, several variants of the so-called SRMR metric were proposed and tested within the scope of the ACE Challenge to predict reverberation time and direct-to-reverberant energy ratio (DRR) parameters. Experiments with single-channel data showed the benefits of the proposed normalization strategy (i.e., NSRMR) to reduce estimator RMSE. A further reformulation of the metric to take into account per-band and per-frame SRMRs led to further gains in RMSE, but more importantly in significant increases in correlation (as much as 47\%) with ground truth data. In all cases, improvements were seen over a state-of-the-art benchmark algorithm. The proposed DRR estimators also showed significant improvements over the original SRMR metric, with reductions of up to 18\% being observed. For multi-channel data, in turn, a simple per-channel feature averaging approach was used and showed comparable results with single-channel data, but with significantly lower prediction error variability (around 50\%) in the 2-channel case.
\section{Acknowledgements}
The authors acknowledge funding from NSERC, FRQNT, and the Nuance Foundation.

 \clearpage
  \balance
\bibliographystyle{IEEEtran}

\bibliography{Fusion_short_long_term}

\end{document}